\documentclass[journal,10pt]{IEEEtran}
\usepackage{amsmath,cite,amsfonts,amssymb,color, mathtools,setspace,comment,float,caption,graphicx}

\def \bmats{\left[\begin{smallmatrix}}
\def \emats{\end{smallmatrix}\right]}\def \bmats{\left[\begin{smallmatrix}}
\def \emats{\end{smallmatrix}\right]}
\def \beqi{\begin{IEEEeqnarray}{rcl}\IEEEyesnumber}
\def \eeqi{\end{IEEEeqnarray}}

\def \bmat{\begin{bmatrix}}
\def \emat{\end{bmatrix}}
\def \beq  { \begin{equation} }
\def \eeq { \end{equation} }
\def \beqn{ \begin{eqnarray} }
\def \eeqn{ \end{eqnarray} }
\usepackage{algpseudocode}
\usepackage{multirow}
\usepackage[linesnumbered,ruled,boxed]{algorithm2e}
\algrenewcommand\textproc{}
\let\oldnl\nl
\newcommand{\nonl}{\renewcommand{\nl}{\let\nl\oldnl}}
\newcommand{\algrule}[1][.2pt]{\par\vskip.3\baselineskip\hrule height #1\par\vskip.3\baselineskip}

\begin{document}

\title{Stable Matching for Selection of Intelligent Reflecting Surfaces in Multiuser MISO Systems}
\author{Jawad Mirza,~\IEEEmembership{Senior Member,~IEEE}, Bakhtiar Ali and Muhammad Awais Javed,~\IEEEmembership{Senior Member,~IEEE}
 \thanks{J. Mirza, B. Ali  and M. A. Javed are with the Department of Electrical and Computer Engineering, COMSATS University Islamabad, Islamabad, Pakistan, (Emails: jaydee.mirza@gmail.com, \{bakhtiar$\_$ali, awais.javed\}@comsats.edu.pk).}}

\maketitle

\begin{abstract}
In this letter, we present an intelligent reflecting surface (IRS) selection strategy for multiple IRSs aided multiuser multiple-input single-output (MISO) systems. In particular, we pose the IRS selection problem as a stable matching problem. A two stage user-IRS assignment algorithm is proposed, where the main objective is to carry out a stable user-IRS matching, such that the sum rate of the system is improved. The first stage of the proposed algorithm employs a well-known Gale Shapley matching designed for the stable marriage problem. However, due to interference in multiuser systems, the matching obtained after the first stage may not be stable. To overcome this issue, one-sided (i.e., only IRSs) blocking pairs (BPs) are identified in the second stage of the proposed algorithm, where the BP is a pair of IRSs which are better off after exchanging their partners. Thus, the second stage validates the stable matching in the proposed algorithm. Numerical results show that the proposed assignment achieves better sum rate performance compared to distance-based and random matching algorithms.
\end{abstract}


\IEEEpeerreviewmaketitle

\section{Introduction}\label{intro}
 \IEEEPARstart{I}{ntelligent} reflecting surface (IRS) is an artificial passive surface that consists of large number of low-cost reflecting elements. By introducing phase shifts and/or amplitude variations, IRS can reflect the incident electromagnetic wave towards the specified direction, thus enabling a smart/programmable wireless environment \cite{wu2019towards}. IRS has been envisioned to revolutionize high frequency wireless communication systems particularly when combined with other promising technologies such as massive multiple-input multiple-output (MIMO) and terahertz communications. More concisely, the quality of the MIMO channel link can be improved, i.e., unfavourable propagation conditions can be controlled by judiciously designing the phase shifts of IRS reflecting elements. 
 
Multiple IRSs aided communication systems have shown to provide robust data transmission and wide coverage area \cite{yang2020energy} compared to the single IRS deployment. This motivates us to investigate user-IRS association problem in multiple IRSs aided multiuser multiple-input single-output (MISO) systems.
 There are few studies that deal with the user-IRS association problem in single-input single-output (SISO) systems, however, here we only discuss related studies investigating this problem in MISO systems. A distance based user-IRS association is performed in \cite{9133435,9148640}, where an IRS is assigned to a nearby user. To get rid off the complicated inter-IRSs interference, orthogonal IRS channels are considered in \cite{li2019joint} and the user-IRS assignment is based on a greedy search algorithm.
 
 In this paper, the user-IRS assignment problem in multiuser MISO systems is modeled as a matching problem. We assume that each user can be matched to at most one IRS, resulting in a one-to-one matching problem such as stable marriage. The seminal studies in matching theory demonstrate that there exist at least one stable matching for general preferences in one-to-one games \cite{gu2015matching}. In this study, the user preference is based on local information available i.e., the user rate without interference. Whereas, the base station (BS) controls and manages preferences of IRSs which are based on the user rate with interference, as it is assumed that the BS has perfect knowledge of global channel state information (CSI). With these two-sided preferences, we propose a two stage IRS optimal stable matching algorithm, where the first stage consists of a well-known Gale Shapley matching algorithm and the second stage identifies blocking pairs (BPs) in the current matching. Due to interference, an IRS choice of a user will impact the choices of the other IRSs in the network, therefore, it is important to identify BPs until a stable matching is obtained. Note that in a stably matched association, there is no single user-IRS pair which is better off, if allowed to change their assigned partners.

\section{System Model}
\begin{figure}[t]
\centering
 \resizebox{.85\linewidth}{!}{
\includegraphics[width=\linewidth]{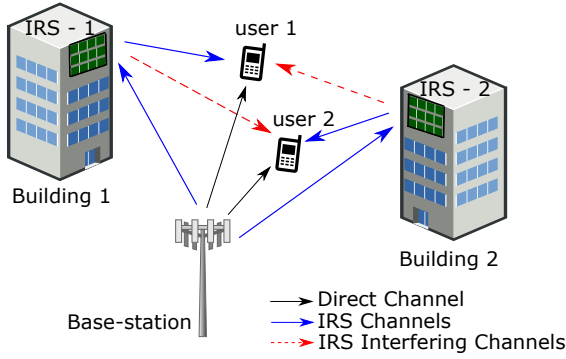}}
\caption{Illustrative system model of the considerd IRSs assisted multiuser MISO system with $K=L=2$.}
\label{fig1}
\end{figure}

Consider a downlink multiuser MISO communication system assisted by multiple IRSs as shown in Figure \ref{fig1}. The system consists of a BS equipped with $M$ transmit antennas.  There are total $K$ number of single-antenna users being served by the BS. In addition to that, there are total $L$ number of IRS units deployed in the surrounding area, where each IRS consists of $N$ reflecting elements.  Let $\mathcal{U}$ and $\mathcal{R}$ denote the set of users and IRSs, respectively. The main objective is to achieve a stable users-IRSs one-to-one matching, $\mu: \mathcal{U} \to \mathcal{R}$, such that the sum rate of the network is improved.   

The baseband equivalent direct channel from the BS to the $k^{\text{th}}$ user, is denoted by $\mathbf{h}_{\text{d},k} \in \mathbb{C}^{M \times 1}$. The channel from the BS to the $l^{\text{th}}$ IRS is represented by $\mathbf{G}_l \in \mathbb{C}^{N \times M}$ and the reflected channel from the $l^{\text{th}}$ IRS towards the user $k$, is denoted by $\mathbf{f}_{k,l} \in \mathbb{C}^{N \times 1}$. In this study, we assume that channels follow quasi-static flat fading, where channel values remain same within a coherence interval $T_c$. Moreover, it is assumed that perfect CSI is available at the BS. The entries of the direct-link channel are assumed to be an independent and identically distributed (i.i.d.) complex Gaussian random variable with zero mean and unit variance, such that, $\mathbf{h}_{\text{d},k}\sim\mathcal{CN}({\bf 0}_M,{\bf I}_M)$. 
Due to close proximity of IRSs and BS, we assume that a line-of-sight (LoS) path exists between the BS and $l^{\text{th}}$ IRS. Therefore, $\mathbf{G}_l$ can be modeled as a Rician fading channel, given by 
\begin{equation}\label{G}
  \mathbf{G}_{l}=\sqrt{\kappa_g/(\kappa_g+1)}\mathbf{G}_{{l}}^{\rm LoS}+\sqrt{1/(\kappa_g+1)}\mathbf{G}_{{l}}^{\rm NLoS},
\end{equation}
where $\kappa_g$ is the Rician factor and $\mathbf{G}_{{l}}^{\rm LoS}\in\mathbb{C}^{N\times M}$ denotes the channel associated with the LoS component, while $\mathbf{G}_{l}^{\rm NLoS}\in\mathbb{C}^{N\times M}$ represents the non-LoS (NLoS) channel matrix, whose entries are i.i.d and follow the complex Gaussian distribution with zero mean and unit variance. 
In \eqref{G}, the fixed LoS channel component ${\bf G}_{l}^{\rm LoS}$ is modeled as\footnote{We use $(\cdot)^H$, $(\cdot)^{*}$, $(\cdot)^T$ and $(\cdot)^{-1}$  to denote the conjugate transpose, the conjugate, the transpose, and the inverse operations, respectively. For any given matrix $\mathbf{A}$, the quantity $\mathbf{A}(i,j)$ denotes the entry of the matrix $\mathbf{A}$ corresponding to the $i^{\text{th}}$ row and $j^{\text{th}}$ column. Similarly, $\mathbf{A}(:,l)$ represents the $l^{\text{th}}$ column of the matrix $\mathbf{A}$. $\mathbf{a}(n)$ denotes the $n^{\text{th}}$ entry of the vector $\mathbf{a}$.}
\begin{equation}
\mathbf{G}_{l}^{\rm LoS}=\mathbf{a}_N\left(\theta_{l}^{\mathrm{AoA}}\right) \mathbf{a}_M^H\left(\theta_{l}^{\mathrm{AoD}}\right),
\end{equation}
where $\theta_{l}^{\mathrm{AoA}}$ and $\theta_{l}^{\mathrm{AoD}}$ represent the angle of arrival (AoA) and angle of departure (AoD) of the $l^{\text{th}}$ IRS, respectively. The $n^{\text{th}}$-dimensional general uniform linear array response vector, denoted by $\mathbf{a}_n(\theta)\in\mathbb{C}^{n \times 1}$, can be expressed as
\begin{equation}
\mathbf{a}_n(\theta)\hspace{-.2em}=\hspace{-.2em}\left[1,e^{j\frac{2{\pi}d}{\lambda}\sin(\theta)},e^{j\frac{4{\pi}d}{\lambda}\sin(\theta)},\ldots,e^{j\frac{2(n-1){\pi}d}{\lambda}\sin(\theta)}\right]^T, \nonumber
\end{equation}
where $\theta$ is the angle, $d$ is the distance between neighbouring elements, and $\lambda$ denotes the wavelength of the carrier. The channel between the $l^{\text{th}}$ IRS and the user $k$ is given by 
\begin{equation}
      \mathbf{f}_{k,l}=\sqrt{\kappa_f/(\kappa_f+1)}\mathbf{f}_{k,l}^{\rm LoS}+\sqrt{1/(\kappa_f+1)}\mathbf{f}_{k,l}^{\rm NLoS},
\end{equation}
where $\kappa_f$ is the Rician factor and $\mathbf{f}_{k,l}^{\rm LoS}=\mathbf{a}_N\left(\theta_{k,l}\right)\in\mathbb{C}^{N \times 1}$ is the fixed LoS channel, where $\theta_{k,l}$ represent the AoD from the $l^{\text{th}}$ IRS to the $k^{\text{th}}$ user. $\mathbf{f}_{k,l}^{\rm NLoS}\in\mathbb{C}^{N\times 1}$ is the NLoS channel vector, whose entries are i.i.d. and follow the complex Gaussian distribution with zero mean and unit variance. These channels are also multiplied by the square root of the distance-dependent path loss, whose general form is given in Section IV. 

Assuming that the $k^{\text{th}}$ user is paired with the IRS $\mu(k)$, where $1 < \mu(k) < L$, then we can define desired and interfering channels at the $k^{\text{th}}$ user as 
\begin{equation}
    \mathbf{h}_{k,\mu(k)}=\mathbf{h}_{\text{d},k}^H + \mathbf{f}_{k,\mu(k)}^H \mathbf{\Phi}_{\mu(k)} \mathbf{G}_{\mu(k)}
\end{equation}
and
\begin{equation}
  \mathbf{z}_{k,l} = \sum_{l\neq\mu(k)}^{L}\mathbf{f}_{k,l}^{H} \mathbf{\Phi}_{l} \mathbf{G}_{l}, 
\end{equation}
respectively, where $\mathbf{\Phi}_{\mu(k)}$ is the diagonal phase shift matrix for the $\mu(k)^{\text{th}}$ IRS, given by $\mathbf{\Phi}_{\mu(k)}=\text{diag}\{\alpha_{1,\mu(k)}e^{j\phi_{1,\mu(k)}},\alpha_{2,\mu(k)}e^{j\phi_{2,\mu(k)}}, \hdots, \alpha_{N,\mu(k)}e^{j\phi_{N,\mu(k)}}\}$, where $\phi_{n,\mu(k)}$ and $\alpha_{n,\mu(k)}$ denote the phase and ON/OFF state of the $n^{\text{th}}$ element of the IRS $\mu(k)$. Here, set of discrete phase shifts are considered for reflecting elements. We can write the received signal at the $k^{\text{th}}$ user assisted by the selected $\mu(k)^{\text{th}}$ IRS as 
\begin{align}\label{rcd_sig}
   & y_{k,\mu(k)} = \sqrt{P_k} \big(\mathbf{h}_{k,\mu(k)} + \sum_{l \neq \mu(k)}^{L} \mathbf{z}_{k,l}\big) \mathbf{w}_{k} s_k + \nonumber \\
    & \sum_{j=1, j\neq k}^{K} \sqrt{P_j} \big(\mathbf{h}_{k,\mu(k)} + \sum_{l \neq \mu(k)}^{L} \mathbf{z}_{k,l}\big) \mathbf{w}_{j} s_j + n_k,
\end{align}
where $P_k$ denotes the transmit power allocated for the $k^{\text{th}}$ user. The transmit precoding matrix for the $k^{\text{th}}$ user at the BS is represented by $\mathbf{w}_k$, where $\| \mathbf{w}_k\|^2 = 1$. The transmitted symbol for the $k^{\text{th}}$ user is given by $s_k$. The additive noise at the $k^{\text{th}}$ user is represented by $n_k$, which is assumed to follow the i.i.d. Gaussian distribution with zero mean and variance $\sigma_k^2$. Using \eqref{rcd_sig}, we can write the SINR of the $k^{\text{th}}$ user as 
\begin{equation}
     \gamma_{k,\mu(k)}= \frac{P_k \left\|\left(\mathbf{h}_{k,\mu(k)} + \sum_{l \neq \mu(k)}^{L} \mathbf{z}_{k,l}\right)\mathbf{w}_k\right\|^2}{\sum_{j\neq k}^{K} P_j \left\|\left(\mathbf{h}_{k,\mu(k)} + \sum_{l \neq \mu(k)}^{L} \mathbf{z}_{k,l}\right)\mathbf{w}_j\right\|^2 + \sigma_k^2}. \nonumber
\end{equation}
By treating multiuser interference as noise, we can express the achievable rate $R_k$ (in bits/s/Hz) at the $k^{\text{th}}$ user as $R_k = \log_2 \left(1 + \gamma_{k,\mu(k)} \right)$. Consequently, the overall sum rate of the network is given by $R_{\text{sum}}=\sum_{k=1}^{K} R_k$. Similar to \cite{8741198}, we employ fixed zero-forcing (ZF) precoding at the BS. The concatenated channel matrix is given by ${\mathbf{H}} = [{\mathbf{h}}_{1,\mu{(1)}}, \hdots, {\mathbf{h}}_{K,\mu{(K)}}]^{T}$. The ZF precoding vector for the $k^{\textrm{th}}$ user is denoted by ${\mathbf{w}}_{k}$, which is the $k^{\text{th}}$ normalized column of the matrix ${\mathbf{W}}$, where ${\mathbf{W}} = {\mathbf{H}}^{H} ( {\mathbf{H}} {\mathbf{H}}^{H})^{-1}$. For the phase shifts or passive beamforming design, we employ an instantaneous SNR maximization approach \cite{9148640} for the reflective link. For the $k^{\text{th}}$ user, the SNR maximization problem is 
\begin{equation}\label{P1}
    \max_{\mathbf{\Phi}_{\mu(k)}}  \ \ \frac{\|\mathbf{f}_{k,\mu(k)}^H \mathbf{\Phi}_{\mu(k)} \mathbf{G}_{\mu(k)}\|^2}{N_0}, \ \ \text{s.t.} \ |\phi_{n,\mu(k)}| = 1, \ \forall n 
\end{equation}
The sub-optimal solution of the problem \eqref{P1} is provided in Algorithm \ref{a.1} \cite{9148640}, where user and IRS indices are ignored for simplicity. It is based on a discrete reflection phase set $\mathcal{Z}=\{-\pi, -\pi+(2\pi/2^B), \hdots, -\pi+(2\pi/2^B)(2^B-1)\}$, where $B$ is the number of IRS control bits. In steps 4 and 5, we have $\Gamma_{n,\hat m}=|\mathbf{f}(n)| |\mathbf{G}(n,\hat m)|$. Although, Algorithm \ref{a.1} provides a sub-optimal solution, but it exhibit low complexity, i.e., $\mathcal{O}(N)$. 
\begin{algorithm}[!t]
\caption{Passive Beamforming Design \cite{9148640}}
\label{a.1}
\DontPrintSemicolon{
{\textbf{Input:}} $\mathbf{G}$, $\mathbf{f}$ and $\mathcal{Z}$\;
{\textbf{Initialize:}} Set $s_{0}=0$, \ \ \ \ \ \ \ \ \ \ \ \ \ \ \ \ \ \ \ \ \ \ \ \ \ \ \ \ \ \ \ \ \ \ \ \ \ \ \ \ \ \ \ 
Select $\hat m = \mathop {\arg \max }\limits_{1 \le m \le M}  \left\| {{{\mathbf{G}}(:,m)}} \right\|$\; 
\For { $n = 1, \hdots, N$,} { 
\textbf{find} $\hat{\theta} = \mathop {\arg \max }\limits_{{\theta} \in \mathcal{Z}} \left| {{s_{{{n - 1}}}} + \Gamma_{n,\hat m} e^{j(\angle\mathbf{f}^{*}(n)+\angle\mathbf{G}(n,\hat m)+ {\theta})}}\right|$\;
\textbf{set} ${s_n} ={s_{n-1}} +  \Gamma_{n,\hat m} e^{j(\angle\mathbf{f}^{*}(n)+\angle\mathbf{G}(n,\hat m)+ {\hat{\theta}})} $\;
\textbf{set} ${\phi _{{n}}} =  \hat{\theta}$\;
}
\textbf{Output:} ${{\boldsymbol{\Phi }}} = {\rm{diag}}\left\{ {e^{j\phi _1},e^{j\phi _2}, \cdots ,e^{j\phi _N}} \right\}$\;
}
\end{algorithm}
\section{Proposed IRS Selection Strategy}
In this section, we present our proposed IRS-user assignment framework for the system model discussed above. Here, we explain three important stages of the selection strategy, namely; CSI acquisition, preference list setup and stable matching. The proposed algorithm is designed for the case when the number of users in the system is equal to the number of IRSs, i.e., $K=L$. Let $\mathcal{U}=\{u_1,u_2,\hdots,u_K\}$ and $\mathcal{R}=\{r_1,r_2,\hdots,r_L\}$ denote the set of users and IRSs, respectively. The aim of the proposed method is to obtain a stable user-IRS matching $\mu : \mathcal{U} \to \mathcal{R}$ that maximize the overall sum-rate of the network, such that
\begin{equation}
    \max_{\mu} \ \sum_{k}^{K} \log_2 \left(1 + \gamma_{k,\mu(k)} \right), \ \text{s.t.} \ \mu \ \text{is a matching}, 
\end{equation}
where $\mu(k)$ denotes the index of the IRS which is matched to the $k^{\text{th}}$ user. Each user will be matched with only one IRS.

{\bf CSI Acquisition:} Although, we have assumed perfect CSI in this study, for practical implementation of the proposed algorithm, it is important to provide the details of CSI acquisition process at the BS and users.

{\emph{a) Global CSI at BS:}} We classify the CSI acquisition stage in two main categories; direct channel and reflected IRS channel estimations. During direct channel estimation, we assume that all IRS units are turned OFF. Here, conventional multiuser MIMO training based TDD  channel estimation technique can be employed by leveraging uplink/downlink channel reciprocity. For the estimation of IRS-based reflected channels, each IRS is turned ON one-by-one, i.e., for the $i^{\text{th}}$ IRS, we have $\alpha_{n,i}=1$, $\forall n$, whereas, $\alpha_{n,j}=0$ where $j \neq i$. When the $i^{\text{th}}$ IRS is turned ON, each user sends an orthogonal training sequence of length $T$ (where $T < T_c$) to the BS in the same time-frequency resource. The BS estimates downlink channels from the received observation using appropriate criteria. For practical channel estimation, we refer the reader to a single IRS based multiuser channel estimation in \cite{liu2020matrix}. CSI acquisition continues at the BS until all the IRS are turned ON and OFF one-by-one in a systematical manner. At the end of the training process, the BS acquires all the downlink user channels, i.e., both direct and reflected channels. 

{\emph{b) Local CSI at users:}} In this study, we assume that the user only have a knowledge of its own channel. For this purpose, downlink training can be used to acquire the local CSI, which is given by $\mathbf{h}_{k,l}$ for the $k^{\text{th}}$ user assisted by the $l^{\text{th}}$ IRS. 

{\bf Preference List Setup:} After the completion of uplink and downlink CSI acquisition process, each user generates its preference list based on the offered rates from IRSs. The channel gain for the $k^{\text{th}}$ user when served by the $l^{\text{th}}$ IRS is given by $|h_{k,l}|^2$. The user computes the rate by using $C_{k,l}=\log_2(1+(|h_{k,l}|^2/\sigma_k^2))$ $\forall \ l$. At the user, the preference list consists of IRSs, which are ranked in a descending order based on their offered  rate. This means that the user's preference list is based on the local channel information (without interference) as users do not have the information of other user channels. The preference list created by the $k^{\text{th}}$ user is denoted by ${Pl\_u}_{k}$. 

On the other hand, the BS has perfect knowledge of global CSI, and therefore, unlike the user preference list, the preference lists created at the BS for IRSs are based on the users rates with interference, i.e., for the $k^{\text{th}}$ user $R_{k,\mu(k)}=\log_2(1+\gamma_{k,\mu(k)})$. This rate can also be expressed with respect to the $l^{\text{th}}$ IRS as $R_{\mu(l),l}=\log_2(1+\gamma_{\mu(l),l})$, where $\mu(l)$ denotes the index of the user which is matched to the $l^{\text{th}}$ IRS. It is not feasible for the BS to compute all the possible user-IRS permutations as this will increase the computational overhead significantly at the BS. Therefore, to compute preference lists at the BS for IRSs, a computationally less complex strategy is designed where the BS finds a small number of random user-IRS permutations, however, it is assured that each user is assigned with all the IRSs. An example for the $K=L=3$ case is presented here to explain the strategy. The BS generates a user-IRS association matrix consisting of total $L$ number of rows instead of $L$! rows if all permutations are considered. This random user-IRS association matrix for the $K=L=3$ case is given by
\begin{equation}
    \mathbf{A} = \begin{bmatrix}
(u_1,r_1) & (u_2,r_2)  & (u_3,r_3)\\
(u_1,r_3) & (u_2,r_1)  & (u_3,r_2)\\
(u_1,r_2) & (u_2,r_3)  & (u_3,r_1)
\end{bmatrix}.
\end{equation}
The rows of the matrix $\mathbf{A}$ define three different user-IRS associations for computing the preference lists of IRSs at the BS. In the first row, the first user is paired with the first IRS, i.e., $u_1 \leftrightarrow r_1$, while other users have $u_2 \leftrightarrow r_2$ and $u_3 \leftrightarrow r_2$ associations. The preference list of the first IRS will computed by sorting the rates of the users with associations $\{\mathbf{A}(1,1), \mathbf{A}(2,2), \mathbf{A}(3,3)\}$ in the descending order. The preference list for the $l^{\text{th}}$ IRS is represented by ${Pl\_r}_{l}$. 
\begin{algorithm}[!htp]
  \caption{User-IRS Assignment Algorithm}
  \label{algo:1}
  \DontPrintSemicolon{
  \KwIn{Set of all users $\mathcal{U}$ and IRSs $\mathcal{R}$,
  user preference lists ${Pl\_u}_{k}$ $\forall$ $k$, IRS preference lists ${Pl\_r}_{l}$ $\forall$ $l$} 
  \algrule
  \nonl \hspace{+.5em} {\emph{Stage 1: Gale-Shapley}}\;
  \algrule
  {\bf{Initialize}} Each IRS $r_l \in \mathcal{R}$ to be free, $\mu = \emptyset$\;
\While {IRS $r_l \in \mathcal{R}$ is free and ${Pl\_r}_{l} \neq \emptyset$}{
$u_k =$ first user on $r_l$'s list to whom $r_l$ has not proposed yet\;
{\bf if} {($u_k$ is not assigned)}\;
{
Assign $u_k$ and $r_l$ to be allocated to each other\;
$\mu \leftarrow \mu \cup (u_k,r_l) $\;
}

{\bf else if} {($u_k$ prefers $r_l$ over previously assigned $r_j$)}\;
{
Assign $r_j$ to be free $\mu \leftarrow \mu/(u_k,r_j) $\;
Assign $u_k$ and $r_l$ to be allocated to each other $\mu \leftarrow \mu \cup (u_k,r_l) $\;
{\bf else}\;
$u_k$ rejects $r_l$ and ($r_l$ remains unassigned)\;
{\bf end if}
}
}
{\bf Output} $\mu$: matched user-IRS pairs\;
\algrule
 \nonl  \hspace{+.5em}{\emph{Stage 2: Stable Matching (identifying BPs)}}\;
 \algrule
Set $\mu_t=\mu$ and $t=0$\;
\While {$\mu_t$ is not Pareto optimal}{
{\bf for} {all IRSs pairs $(r_i,r_j)$ {\bf do}}\;
{
Switch users of pair:\hspace{-.2em} $(\mu_t(r_j),r_i)$ \hspace{-.2em}and\hspace{-.2em} $(\mu_t(r_i),r_j)$\;
Compute new user rates $R_{\mu_t(j),i}$ and $R_{\mu_t(i),j}$\;
{\bf if} {($R_{\mu_t(j),i} \hspace{-.1em}>\hspace{-.1em}R_{\mu_t(i),i}$ \hspace{-.1em}and\hspace{-.1em} $R_{\mu_t(i),j}\hspace{-.1em}>\hspace{-.1em}R_{\mu_t(j),j}$)}\;
{
IRS pair $(r_i,r_j)$ is allowed to exchange users\;
{\bf else}\;
IRS pair $(r_i,r_j)$ is not allowed to exchange users\;
{\bf end if}\;
}
{\bf end for}\;
}
Set $t=t+1$\;
}
{\bf Output} $\mu_s=\mu_t$: stably matched user-IRS pairs\;
}
\end{algorithm}

{\bf Stable Matching:} The studied problem is a bipartite matching problem with two-sided preferences. The proposed user-IRS assignment algorithm comprises of two phases: 1) Gale-Shapley and 2) blocking pair (BP) identification for stability. The pseudo code of the proposed algorithm is given in Algorithm \ref{algo:1}. Each user shares its preference list with the BS. Although, the BS has global CSI available and it can compute users preference lists, however, due to imperfect uplink/downlink channel reciprocity in practical systems, we rely on users to compute and share their preferences.

After obtaining the preference list of the users, the BS performs the proposed user-IRS assignment which is IRS optimal. For the given IRS, the BS assigns the most preferred user to the IRS, if it is not already matched with any other IRS. If that preferred user is already matched to one of the other IRSs, then it is re-assigned to the proposing IRS only if the user also prefers it over the assigned IRS. The same process is repeated for all the IRSs until all the IRSs are matched. 

Since, the IRS matching with any given user will also effects the performance of other users, therefore the matching output $\mu$ at step 14 may not be stable. This instability is due to the interference which could be caused by IRSs assigned to other users. The structure of ZF precoding also effects the stability at this phase. The matching obtained with such interdependence is called as matching games with externality \cite{gu2015matching}. Therefore, in the second stage of the Algorithm \ref{algo:1}, the BS finds the BPs in the current matching $\mu$. A BP is a pair of user  
and IRS $(u,r)$, who prefers each other over their current partners. The BS searches for all unstable IRSs pairs such that if the rate obtained by exchanging users is beneficial for both the IRSs then exchange is allowed. This one-sided (only IRSs) stability is called Pareto optimality in matching theory, where there is no other matching in which some IRS is better off, while no IRS is worse off. The process continues until a trade-in-free environment is reached, resulting in a stable matching $\mu_s$.

\section{Simulation Results}
In this section, we run extensive simulations to access the performance of our proposed user-IRS matching algorithm. We consider a cellular communication setup where a single BS is located at the origin $(0,0)$ and IRSs are distributed equispaced on a circle around the BS with radius $d_{R}$. Users are deployed uniformly at each realization inside the circle with maximum spread equal to $d_{R}/2$ from the BS. Throughout the simulations, the value of distance is set to $d_{R}=50$, unless stated otherwise. All simulation results are obtained by statistically averaging over large number of channel realizations. 

The path loss between the BS-IRS and IRS-user links are modelled as $L_{bi}=K_{bi}({d_{bi}})^{-\alpha_2}=C_\nu\delta_b\delta_i({d_{bi}})^{-\alpha_2}$ and $L_{iu}=K_{iu}({d_{iu}})^{-\alpha_2}=C_\nu\delta_i\delta_u({d_{iu}})^{-\alpha_2}$, respectively, where $\alpha_2=2$ is the path loss exponent. $d_{bi}$ and $d_{iu}$ are the distances of the BS to IRS and IRS to user, respectively. The quantities $\delta_b$ and $\delta_u$ are antenna gains of the BS and user antennas respectively, while $\delta_i$ is the reflection gain of the IRS. The path loss of the IRS assisted composite link can be written as 
\begin{equation}
    L_{biu}=L_{bi}L_{iu}=C_\nu^2\delta_b\delta_u\delta_i^2({d_{bi}d_{iu}})^{-\alpha_2},
    \label{PLIRS}
\end{equation}
The relative reflection gain is given by $\zeta=\delta_i/(\sqrt{\delta_b\delta_u})\Rightarrow \delta_i^2=\zeta^2\delta_b\delta_u$. We can also write \eqref{PLIRS} as $L_{biu}=(C_\nu)^2\zeta^2({d_{bi}d_{iu}})^{-\alpha_2}$,
where we have kept $\zeta=10$~dB. For the direct link between the BS and user, the path loss exponent is taken as $\alpha=3.5$. Throughout this section, we set $\kappa_g=\kappa_f=10$, $C_\nu=-30$ dB and $\delta_b=\delta_u=0$ dB.

\begin{figure}[t]
\centering
\includegraphics[width=\linewidth]{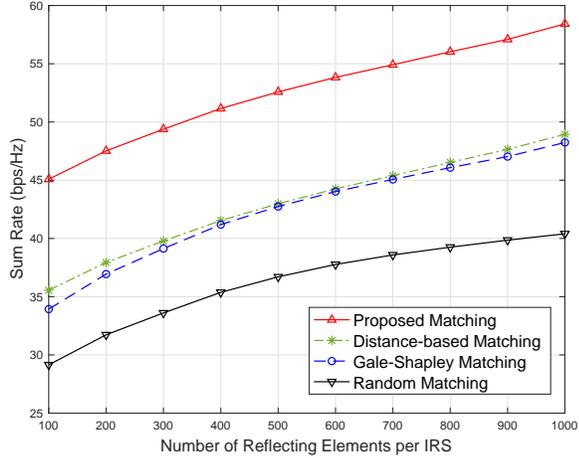}
\caption{Sum rate versus number of Reflecting elements.}
\label{fig2}
\end{figure}

The impact of increasing IRS reflecting elements on the sum rate performance is shown in Fig.~\ref{fig2} with $L=K=M=10$. The total transmit power is kept at $10$ dB, which is equally distributed among the users. For comparison, we also plot sum rate results for distance-based matching \cite{9133435,9148640}, original Gale Shapley matching (i.e., only phase 1 of the Algorithm \ref{algo:1}) and random matching. It can be seen from Fig.~\ref{fig2} that the proposed user-IRS assignment algorithm provides the best sum rate performance as compared to the other schemes. The performance of the standard Gale Shapely matching is comparable with the performance of the distance based matching algorithm where the IRSs are assigned to the nearest user. The distance-based matching is performed using accurate distances, which is not possible in practical systems.

\begin{figure}[t]
\centering
\includegraphics[width=\linewidth]{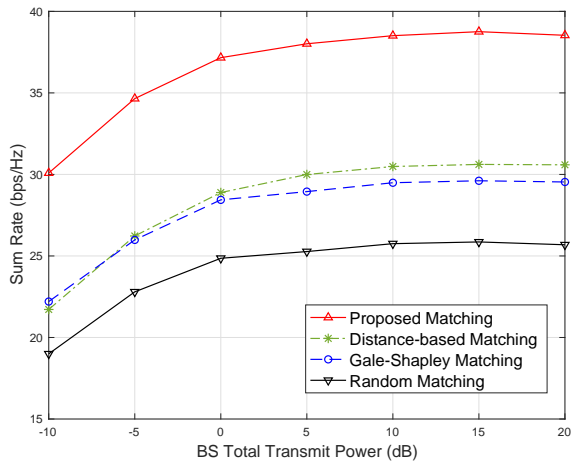}
\caption{Sum rate versus BS total transmit power.}
\label{fig3}
\end{figure}

Fig.~\ref{fig3} shows the sum rate performance of the proposed algorithm for various values of the total transmit power at the BS. Here, we use $L=K=M=8$ and $N=50$. From Fig.~\ref{fig3}, it is noticed that the increase in sum rate is more evident in low transmit power regimes. The reason for this trend is that at high transmit powers interference also rises in the network. Among the schemes plotted, the proposed algorithm has the superior sum rate performance. 

The effect of increasing the deployment radius $d_{R}$ on the achievable sum-rate is captured in Table \ref{T1}. The parameters kept are similar to that of Fig.~\ref{fig3}, except for the value of transmit power which in this case is kept at $9$ dB. The sum rate initially improves as inter-user interference reduces as $d_{R}$ increases. However, as $d_R$ increases further, the higher path loss results in a performance degradation. This suggests that more reflective elements are needed to overcome this degradation. 
\begin{table}[t]
\caption{Sum rate (bps/Hz) results for different $d_R$ values} 
\centering 
\label{T1}
\begin{tabular}{c c c c c c} 
\hline\hline 
\multirow{2}{*}{Method} &  \multicolumn{4}{c}{$d_R$}   \\
 \cline{2-6}
 & 50 & 100  & 150 & 200 & 250  \\ [0.5ex] 
\hline 
Proposed Matching & 37.0 & 38.3 & 38.6 & 38 & 35.8\\ 
Distance Matching & 29.2 & 29.9 & 29.6 & 29.1 & 27.0\\ [1ex] 
\hline 
\end{tabular}
\label{table:nonlin} 
\end{table}

\begin{figure}[t]
\centering
\includegraphics[width=\linewidth]{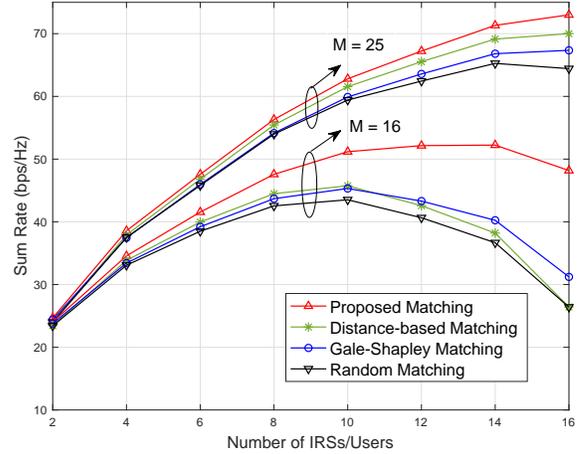}
\caption{Sum rate versus number of IRSs/users $(K=L)$.}
\label{fig5}
\end{figure}

Finally, Fig.~\ref{fig5} shows the sum rate performance against different numbers of IRSs/users in the network. Here, we have $N=20$ and the total transmit power at the BS is $5$ dB. The results are plotted for two different values of $M$, i.e., $M=16$~and~$M=25$. Intuitively, the sum rate increases as the number of IRSs/users increases. It has been noticed that the performance with $M=16$ starts degrading as $K\to M$, suggesting that more transmit antennas or reflecting elements are required to maintain the performance gain. 

\section{Conclusion}
We have proposed an IRSs selection strategy based on stable matching for multiuser MISO communication systems. To achieve user-IRS association in the network that improves the overall sum rate, we rely on two stage matching algorithm. In the first stage the Gale Shapley algorithm is used to find the user-IRS matching. In the second stage of the algorithm, IRSs BPs are determined who are willing to exchange their users, if it is beneficial for both IRSs. Through simulations, it is revealed that the proposed user-IRS assignment outperforms distance-based and random matching algorithms. For future work, it will be useful to investigate user-IRS assignment for MIMO systems by jointly optimizing passive and active beamforming. 
\bibliographystyle{IEEEtran}
\bibliography{ref}

\end{document}